\documentclass{article}
\PassOptionsToPackage{numbers,sort&compress}{natbib}
\usepackage[preprint]{neurips_2025}

\usepackage[T1]{fontenc}
\usepackage[utf8]{inputenc}

\usepackage{amsmath, amssymb}
\usepackage{graphicx}
\usepackage{float}
\usepackage{caption}
\usepackage{listings}
\usepackage[most]{tcolorbox}
\usepackage{booktabs}
\usepackage{mathtools}
\usepackage{url}
\usepackage{tikz}
\usetikzlibrary{arrows.meta,calc}

\usepackage{listings}

\usepackage{natbib}
\title{NISPO: Open-source IUPAC name generation tool}

\author{
  Nicholas T. Runcie \\
  Department of Statistics \\
  University of Oxford \\
  Oxford, UK \\
  \texttt{runcie@stats.ox.ac.uk} \\
  \And
  Fergus Imrie \\
  Department of Statistics \\
  University of Oxford \\
  Oxford, UK \\
  \texttt{imrie@stats.ox.ac.uk} \\
  \And
  Charlotte M. Deane \\
  Department of Statistics \\
  University of Oxford \\
  Oxford, UK \\
  \texttt{deane@stats.ox.ac.uk} \\
}

\begin{document}

\maketitle

\begin{abstract}
Systematic International Union of Pure and Applied Chemistry (IUPAC) names are standard for communicating molecular structures in chemical literature, patents, and databases.
We introduce NISPO, an open-source RDKit-based Python package for IUPAC name generation.
NISPO was developed by an agentic self-improvement loop using OpenAI's Codex with the GPT-5.5 model. 
A generated name was considered correct if the open-source OPSIN tool parsed it back to the input structure.
Guided by this objective, the agent implemented and refined NISPO against 2.68 million molecules from SureChEMBL, resulting in the tool achieving 98.1\% round-trip accuracy on a held-out set of 103 million PubChem molecules.
NISPO is freely available at https://github.com/oxpig/nispo.
\end{abstract}

\section{Introduction}

Chemical names are a core method used for communicating molecular structures. 
A widely used nomenclature for naming molecules was established by the International Union of Pure and Applied Chemistry (IUPAC) and is specified in the IUPAC Blue Book \cite{favre_nomenclature_2014}.
These names are widely used throughout chemistry, for example in academic literature, in patents, and within vendor databases. 
Additionally, given IUPAC names act as a natural language description of molecules, it has been proposed that IUPAC names could be used as a molecular representation within Large Language Models (LLMs) for chemistry \cite{mao_iupac-gpt_2026}. 
Given the widespread use of IUPAC names, a software solution for writing these names would be valuable \cite{williams_need_1999}.

Numerous proprietary tools exist for writing IUPAC names of molecules, such as OpenEye's Lexichem \cite{openeye_lexichem_2025,chemaxon_naming_toolkit_2026,acdlabs_acdname_2025,revvity_chemdraw_2026,dassault_biovia_draw_2026}. 
Previous work proposed training a machine learning model to translate SMILES strings \cite{weininger_smiles_1988} to IUPAC names \cite{rajan_stout_2024}; however, this tool is no longer publicly available. 
Alternatively, LLMs such as ChatGPT \cite{OpenAI_gpt5_system_card} have been used for IUPAC name generation; however, these systems are unreliable and resource-intensive, with GPT-5 achieving only 43.5\% accuracy on SMILES-to-IUPAC translation in the MolJSON benchmark \cite{runcie2026MolJSON}. 
Within the past two weeks, OpenClatura was released as a preprint describing an open-source tool for generating IUPAC names \cite{OpenClatura}.

The Open Parser for Systematic IUPAC Nomenclature (OPSIN) provides an established open-source solution for the reverse task of converting IUPAC names into molecular structures \cite{lowe_chemical_2011}.
In its original publication, OPSIN reported 99.8\% precision and 98.7--99.2\% recall at converting IUPAC names to molecular structures, and the tool has been actively maintained since its release.
OPSIN is now widely used for name-to-structure conversion \cite{o_boyle_2025_16755947}, including within LLM research \cite{m_bran_augmenting_2024,sharlin_nmr-challenge_2026,ChemIQ,runcie2026MolJSON}.

In recent months, coding agents such as Codex \cite{openai_introducing_codex_2025} and Claude Code \cite{anthropic_claude_code_2025} have emerged as powerful tools for software development \cite{yang_swe-agent_2024}.
These systems provide an LLM with tools that allow it to inspect codebases, edit files, run tests, and iterate on code.
Coding agents are rapidly improving and can now work autonomously on tasks for hours or days \cite{kwa2026measuringaiabilitycomplete}.
A variety of orchestration strategies have been explored for running agents autonomously; an example of this is the ``Ralph loop'' in which agents repeatedly prompt themselves until a specified objective is achieved \cite{huntley_ralph_2025,huntley_everything_2026}.
Such a method is particularly effective when progress can be measured by a verifiable objective, such as accuracy on a benchmark.
Verifier-guided agentic approaches have shown promise in multiple fields, such as AlphaEvolve for novel algorithm discovery \cite{novikov2025alphaevolvecodingagentscientific}, AutoResearch for autonomous improvement of LLMs \cite{karpathy_autoresearch_2026}, and ReactionClassifier, which is an autonomously generated reaction classification taxonomy \cite{armstrong2026agenticgenerationverifiablerules}. 

We introduce and describe the development of NISPO, an open-source RDKit-based Python package for generating IUPAC names of molecules. 
NISPO was created through self-improvement loops using OpenAI's Codex with the GPT-5.5 model. 
The development of NISPO was semi-autonomous, with the human developer defining objectives and providing steering prompts to guide the agent towards the final implementation. 
NISPO was optimised against $\sim$2.68 million molecules from SureChEMBL \cite{papadatos_surechembl_2016,o_boyle_2025_16755947} and evaluated on a held-out set of 103 million molecules from PubChem \cite{kim_pubchem_2025}, achieving 98.1\% round-trip accuracy.
NISPO is among the first examples of scientific software developed through an agentic self-improvement loop.
In this paper, we describe the loop framework used to develop NISPO, show how its performance improved over time, and provide a qualitative analysis of the names it generates. 
NISPO is freely available on our GitHub repository \mbox{\url{https://github.com/oxpig/nispo}}, and can be installed from the Python Package Index using \texttt{pip install nispo}.

\section{Method}

\begin{figure}
    \centering
    \includegraphics[width=1\linewidth]{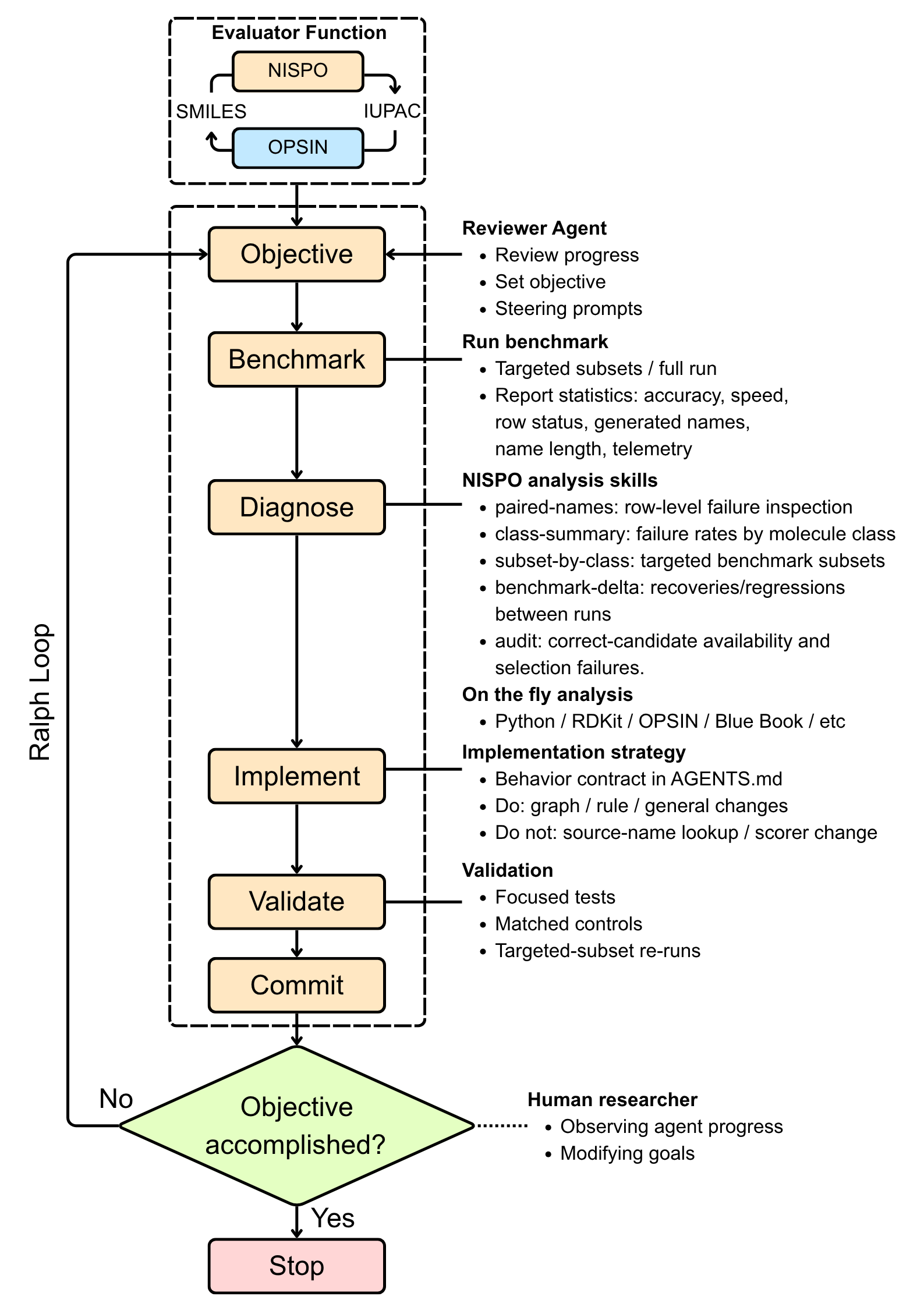}
    \caption{Self-improvement loop used to implement NISPO. The evaluator function defines the correctness of a generated name by an OPSIN round-trip check. The agent was tasked to improve NISPO towards a specified objective, with a Ralph loop repeatedly re-prompting the agent until the objective was achieved. The right column shows skills and context available to the agent for each stage. The sequence of steps reflects the general workflow self-reported by Codex; the agent was free to choose its own strategy, and other sequences may also have been used.}
    \label{fig:implementation_loop}
\end{figure}

\subsection{Overview}
NISPO is an open-source, RDKit-based Python package for generating IUPAC names of molecules.
A generated name is defined as correct if OPSIN \cite{lowe_chemical_2011} parses it to a structure identical to the input molecule. 
This round-trip check provides a deterministic verifier for name correctness, enabling NISPO to be developed through an agentic self-improvement loop. 
Using Codex with the GPT-5.5 model, the agent iteratively improved NISPO to maximise round-trip accuracy on 2.68 million molecules from the SureChEMBL dataset \cite{papadatos_surechembl_2016,o_boyle_2025_16755947}; the resulting tool was then evaluated on a held-out set of 103 million PubChem molecules \cite{kim_pubchem_2025} and achieved 98.1\% accuracy. 
All NISPO code was written by the coding agent, with the human researcher providing optimisation objectives and steering prompts. 
The remainder of this section describes the datasets, objectives, and loop strategy used to develop NISPO.

\subsection{Task definition and success criteria}
\label{sec:task_definition_and_success_criteria}

The objective of this work was to develop a tool for writing IUPAC names of molecules. 
Performance was measured by a round-trip check with OPSIN: names generated by NISPO were converted to SMILES strings by OPSIN (version 2.9.0) \cite{lowe_chemical_2011}, and the isomeric canonical SMILES of the input and OPSIN-parsed output structures, generated by RDKit (version 2026.03.3) \cite{rdkit}, were compared for exact match.
This check is deterministic, providing a verifiable objective function for the coding agent to optimise. 
The evaluation function was implemented as a Python script that ran NISPO and performed the round-trip check. 
In addition to accuracy, the evaluator recorded diagnostic information for each run, including runtime statistics, failure-stage analysis, and telemetry describing internal function usage within NISPO. 
These diagnostics were not part of the optimisation objective, but the implementation agent used them for debugging and for prioritising subsequent modifications.

Two further objectives were introduced during development. 
Early NISPO versions produced long, nested names with poor parent-structure selection; to mitigate this, we introduced an objective to minimise the mean name length. 
We also found that NISPO became slower as accuracy improved, so we added a third objective of minimising the average runtime per molecule. 

\subsection{Datasets}

NISPO was developed using molecules from the SureChEMBL dataset, which contains approximately 4 million molecules with IUPAC names extracted from patent literature \cite{o_boyle_2025_16755947,papadatos_surechembl_2016}. 
We applied a prefilter, selecting molecules with at most 50 heavy atoms, elements in the set \{H, C, N, O, S, P, F, Cl, Br, I\}, a single fragment, at least one carbon atom, no radical electrons, and no isotopic labels; no filter was applied for stereochemistry or formal charge. 
This yielded a set of approximately 2.68 million molecules. 

Development followed a curriculum strategy. 
The initial NISPO tool was bootstrapped using 9,143 molecules from the MolJSON benchmark \cite{runcie2026MolJSON}. 
A random sample of 50,000 SureChEMBL molecules was then chosen as a starting benchmark. 
Once accuracy on the current subset reached a sufficient level, as judged by the human researcher, a new subset was constructed from a sample of the remaining failing molecules. 
Across the full NISPO implementation, 18 benchmark subsets were created in this way, repeatedly focusing the agent on remaining failure modes. 

The final NISPO tool was evaluated using a set of approximately 103 million molecules selected from PubChem \cite{kim_pubchem_2025} using the same criteria as above.
Approximately 87.9\% of the SureChEMBL set is also present in the PubChem test set, corresponding to $\sim$2.27\% of the test set.
The PubChem benchmark was run by the human researcher at intervals during NISPO development to observe general progress; the results of the PubChem benchmark were not used by the agent for optimisation. 
This benchmark was run on the University of Oxford Advanced Research Computing high-throughput cluster \cite{oxford_arc}, with each 100,000-molecule naming shard executed on a single core of an AMD EPYC 9634 processor (Genoa, 2.25 GHz).

\subsection{Agentic self-improvement loop}
\label{sec:agentic-loop}

All implementation was performed by a Codex agent running the GPT-5.5 model \cite{openai_introducing_codex_2025}, referred to herein as the ``implementation'' agent. 
The project description and behaviour contract for the implementation agent were specified in the AGENTS.md file (Appendix Listing \ref{lst:original-agents-file}), which the agent read at the start of each thread.
A second Codex agent, referred to as the ``reviewer'' agent, was used by the human researcher to supervise and steer the implementation loop, e.g. analysing the codebase, modifying the loop framework, and defining new objectives.
The reviewer agent was prompted by the human researcher to generate goal prompts for the implementation agent; an example is shown in Appendix Listing \ref{lst:goal-prompt}.
Each goal prompt was given to Codex GPT-5.5 using the \texttt{/goal} mode, which repeatedly re-prompts the agent until the objective is met, following the Ralph loop strategy \cite{huntley_ralph_2025,huntley_everything_2026}.
The implementation agent followed an approximate cycle of measuring performance, diagnosing failures, and implementing improvements (Figure \ref{fig:implementation_loop}); the specific workflow was not enforced by the harness, with the agent free to choose its own strategy. 

To support development, the agent was provided with a set of reference resources and skills. 
To measure performance, the agent ran a benchmark script, described in Section \ref{sec:task_definition_and_success_criteria}, which reported the objective metrics of accuracy, name length, and runtime, along with per-row analysis and debug information. 
To help analyse the results of benchmark runs, the agent was provided with a set of skills referred to as the ``NISPO analysis suite'' (Appendix Listing \ref{lst:nispo-analysis-suite}).
These included scripts for checking regressions between runs, and for comparing generated NISPO names to reference names in the SureChEMBL set; the agent was also free to write ad hoc analysis scripts for cases not covered by the suite.
Additionally, the graphify skill \cite{shamsi_graphify_2026} was installed, providing the agent with a queryable knowledge graph of the codebase. 
Finally, throughout development, the agent had access to two reference resources: the IUPAC Blue Book, which defines the IUPAC nomenclature rules \cite{favre_nomenclature_2014}, and the OPSIN source repository, which contains name-to-structure conversion logic together with resources such as lists of retained names and parent scaffolds \cite{lowe_chemical_2011}.

Development followed a ``human-on-the-loop'' approach, with the implementation loop running autonomously and the human researcher monitoring progress and intervening when required.
The AGENTS.md file was updated periodically to change the behaviour of the implementation agent; 
for example, a set of ``integrity'' rules was introduced to prevent reward-hacking behaviours, such as compacting output names using string rewrites rather than improving the underlying naming rules (Appendix Listing \ref{lst:final-agents-file}).
The agent often favoured local optimisations over general improvements to the codebase; in these cases, steering prompts were used instructing the agent to ``take a step back and look for step-change improvements''.
When the agent appeared unable to find further improvements, the reviewer agent was used to analyse the codebase and create a detailed improvement plan, as a markdown file, which was given to a fresh implementation agent as context. 
The agent typically struggled with multi-objective goals of the form ``maximise accuracy, minimise name length, and minimise runtime''.
The optimisation was improved by defining upper or lower bounds of performance for two metrics, and prompting the agent to optimise the third. 
The human researcher was required to set interim objectives at each stage in order to guide NISPO towards the desired final properties.

\section{Results and Discussion}

NISPO is an open-source, RDKit-based Python package for generating IUPAC names of molecules. 
It was implemented by Codex using the GPT-5.5 model, through a self-improvement loop in which names were considered correct if OPSIN parsed them back to the input structure. 
In the following sections we report benchmark results on a set of 103 million PubChem molecules (Section \ref{sec:top_level_results}), examine the emergence of naming capabilities and the balancing of multiple objectives during development (Sections \ref{sec:emergence_of_naming_capabilites}, \ref{sec:MOO}), and qualitatively inspect names generated by NISPO (Section \ref{sec:qualitative_name_analysis}).

\subsection{NISPO achieves high round-trip accuracy}
\label{sec:top_level_results}

The agentic self-improvement loop successfully developed NISPO, an IUPAC naming tool with high round-trip accuracy across large and diverse molecular datasets.
NISPO reached a round-trip accuracy of 98.7\% on the SureChEMBL development set of 2.68 million molecules, and 98.1\% on the 103 million molecule PubChem test set, indicating that the naming rules generalise across a large chemical space. 
The IUPAC names listed in PubChem, generated by the proprietary tool Lexichem \cite{openeye_lexichem_2025}, achieved 99.0\% on the same round-trip evaluation. 
This means NISPO approaches the accuracy of established commercial tools, although this comparison only considers OPSIN round-trip success and does not capture adherence to preferred IUPAC nomenclature.

\subsection{Naming capabilities emerged in phases}
\label{sec:emergence_of_naming_capabilites}

\begin{figure}
    \centering
    \includegraphics[width=1\linewidth]{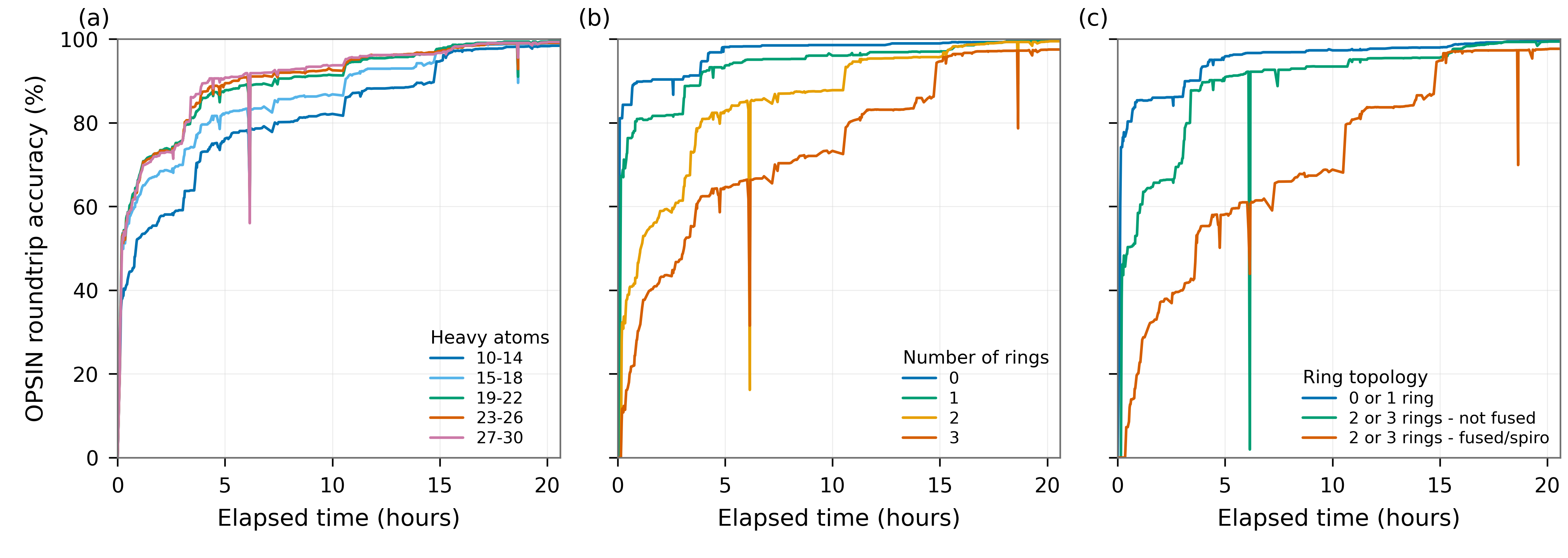}
    \caption{Initial phase of NISPO development optimising accuracy on the MolJSON benchmark set of 9,143 molecules \cite{runcie2026MolJSON}. Performance is stratified by structural class (a) heavy atom count (b) ring count (c) ring topology.}
    \label{fig:moljson-implementation-stratified}
\end{figure}

The initial version of NISPO was implemented by optimising round-trip accuracy against the MolJSON benchmark of 9,143 molecules \cite{runcie2026MolJSON}. 
The accuracy of each benchmark run performed by the agent during this phase is shown in Figure \ref{fig:moljson-implementation-stratified}, stratified by heavy atom count, ring count, and ring topology. 
While NISPO's performance generally improved throughout development, naming capabilities emerged in phases, with structurally simple molecules being supported earlier, and more complex structures requiring more iterations. 
Despite being developed using only 9,143 molecules and in under 20 hours, this version achieved 70.6\% round-trip accuracy on the held-out PubChem test set, indicating the agent had implemented general nomenclature rules (Figure \ref{fig:accuracy_length_speed}a). 

Accuracy improved at a similar rate across different heavy atom counts, with the smaller molecules being slightly more challenging (Figure \ref{fig:moljson-implementation-stratified}a); this is due to the dataset containing a higher proportion of fused and spiro compounds for molecules with fewer heavy atoms \cite{runcie2026MolJSON}. 
There was large variation in improvement rate depending on the ring systems within the molecules.
For molecules with zero rings, 97\% accuracy was achieved in under five hours, whereas molecules with three rings took over 16 hours (Figure \ref{fig:moljson-implementation-stratified}b). 
Furthermore, grouping by ring topology, molecules with fused or spiro ring systems were substantially harder to optimise than molecules without these features (Figure \ref{fig:moljson-implementation-stratified}c). 

The trajectory contains multiple downward spikes in accuracy. 
These reflect cases where the agent implemented a change that harmed NISPO performance, such as a partially implemented feature or a bug. 
In these cases, the agent rectified the issue by either fixing the intended implementation, or reverting to an earlier repository state.

\subsection{Self-improvement loops optimised NISPO across multiple objectives}
\label{sec:MOO}

\begin{figure}
    \centering
    \includegraphics[width=1\linewidth]{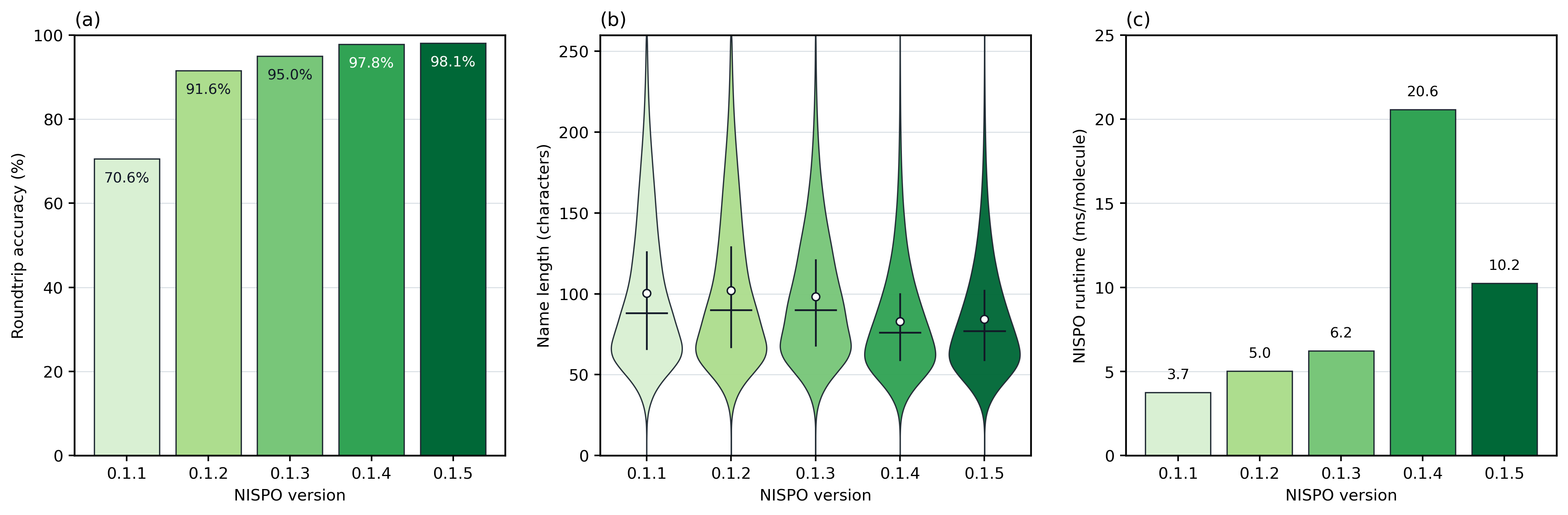}
    \caption{Performance metrics of NISPO versions, benchmarked on the set of 103 million PubChem molecules: (a) round-trip accuracy, (b) mean name length, and (c) mean runtime per molecule.}
    \label{fig:accuracy_length_speed}
\end{figure}

Following the initial bootstrap phase, development continued on the SureChEMBL set using three sequentially introduced objectives: round-trip accuracy, mean name length, and runtime per molecule.
These metrics are plotted in Figure \ref{fig:accuracy_length_speed} for intermediate checkpoints of NISPO.

The initial objective was to optimise round-trip accuracy. 
Following the curriculum strategy outlined in Section \ref{sec:task_definition_and_success_criteria}, accuracy steadily improved across versions 0.1.1 to 0.1.3, reaching a round-trip accuracy of 95.0\% (Figure \ref{fig:accuracy_length_speed}a). 
However, inspection of the names generated by NISPO 0.1.3 revealed they were substantially longer than the reference IUPAC names within PubChem (mean length 98.2 vs 80.9) (Figure \ref{fig:accuracy_length_speed}b). 
This was due to the tool generating exhaustive systematic names rather than using retained names of functional groups and parent scaffolds. 
As a proxy for name quality, name-length was introduced as an additional objective for subsequent NISPO versions, and resulted in the mean name length decreasing to 83.0 characters for NISPO 0.1.4. 
Despite this additional constraint, round-trip accuracy increased 
from 95.0\% to 97.8\% during this phase.

It was observed across development that NISPO became slower as the tool improved (Figure \ref{fig:accuracy_length_speed}c), suggesting the agent was implementing an inefficient solution to the task. 
To control for this, the average run-time of NISPO per molecule was recorded and was used as a third optimisation objective. 
While this metric was originally introduced for NISPO 0.1.2, it was found that strict restrictions on run-time prevent the agent from improving round-trip accuracy. 
Consequently, this metric was used as a ceiling constraint throughout optimisation with the upper bound being slowly relaxed over time. 
By NISPO version 0.1.4 the mean runtime was 20.6 ms per molecule. 
As a final objective, the implementation agent was tasked to reduce run-time while preserving the accuracy and name length metrics. 
The resulting NISPO 0.1.5 achieved a run-time of 10.2 ms per molecule, with a slight improvement in round-trip accuracy. 

In general, the agent struggled to optimise multiple objectives in parallel. 
For example, given the goal: ``maximise accuracy, minimise name length, minimise run-time'', the agent would implement a change but immediately revert it if any metrics regressed on the benchmark, preventing necessary trade-offs between objectives. 
The behaviour was mitigated using the bounded goal strategy described in Section \ref{sec:agentic-loop}, for example maximising accuracy, while keeping name length and runtime below a threshold.

\subsection{NISPO generates valid but non-preferred names}
\label{sec:qualitative_name_analysis}

\begin{figure}
    \centering
    \includegraphics[width=1\linewidth]{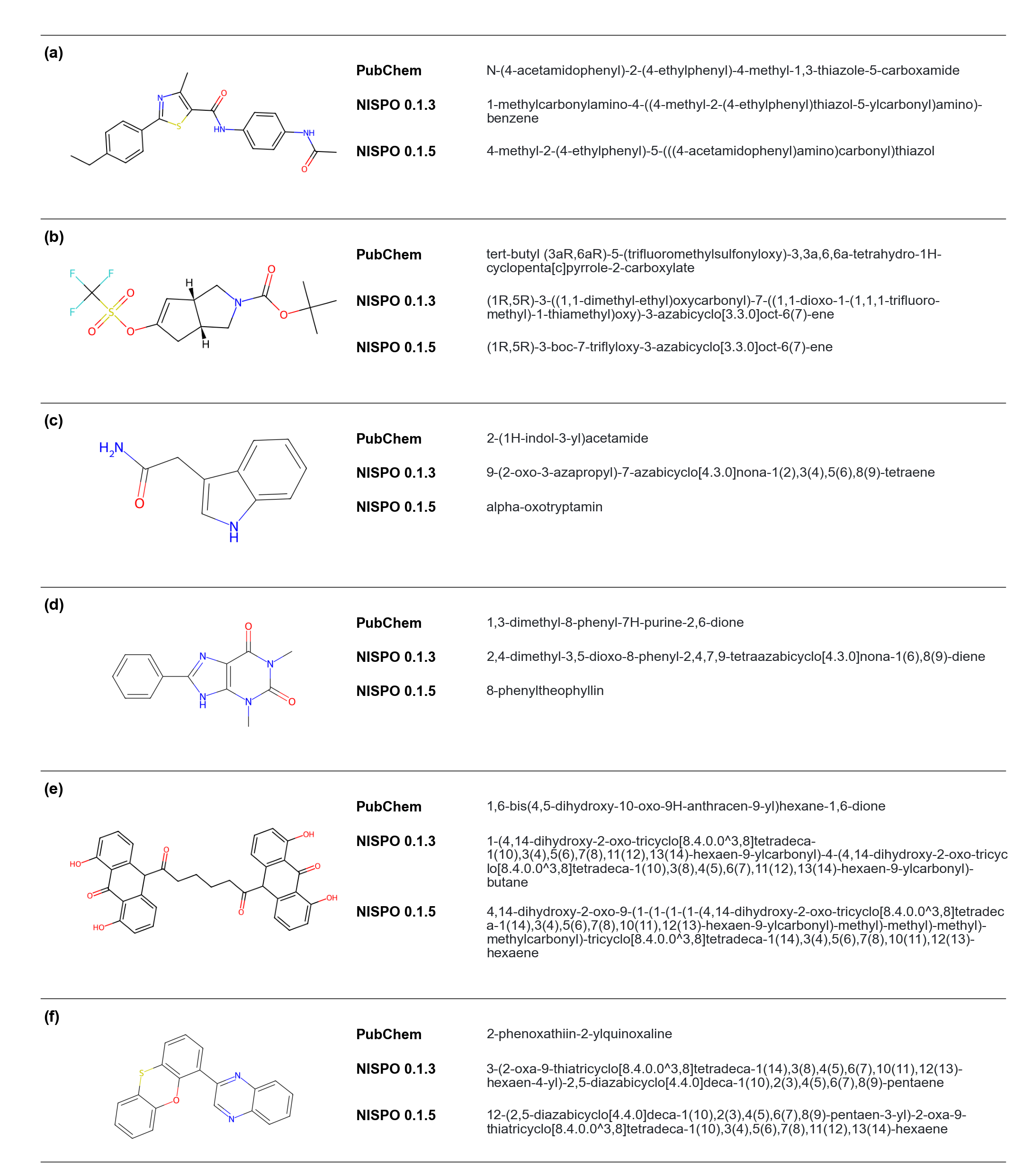}
    \caption{Example names generated by NISPO. The figure shows the IUPAC name listed in PubChem, and the names generated by NISPO version 0.1.3 and NISPO version 0.1.5. All names in this figure pass the OPSIN round-trip test.}
    \label{fig:example_names}
\end{figure}

NISPO is optimised for OPSIN round-trip validity as opposed to preferred IUPAC names; consequently, the names generated by NISPO typically differ from those listed in databases.
Example names are shown in Figure \ref{fig:example_names}, comparing the PubChem-listed IUPAC names with those generated by NISPO versions 0.1.3 and 0.1.5.

While the names generated by NISPO 0.1.3 were valid, they were substantially longer than the reference names in PubChem. 
After the name-length objective was introduced, NISPO 0.1.5 produced shorter names by using more retained names and abbreviated group prefixes.
For example, molecule (b) shows NISPO 0.1.5 using the ``boc'' and ``triflyloxy'' functional group abbreviations, which are supported in OPSIN but are not standard in IUPAC nomenclature. 
Similarly, the choice of parent structure also changed. 
For molecule (c) the tool chose the parent structure ``tryptamine'' and modified this with the ``alpha-oxo'' to form an amide, as opposed to using a more standard naming like ``indole-3-acetamide''. 
NISPO also omitted the final ``e'', yielding ``alpha-oxotryptamin'', which is a truncation tolerated by OPSIN but is not standard for IUPAC nomenclature. 

Despite the name-length objective, the tool can still generate verbose outputs. 
For example, for molecule (e) the NISPO 0.1.5 name contains the highly nested structure ``1-(1-(1-(1-...methyl)methyl)methyl)methyl)'', which could have been replaced with ``butane-1,4-diyl'', and molecule (f) is written in full von Baeyer nomenclature rather than using the retained names quinoxaline and phenoxathiine.

\section{Conclusion}
We have introduced NISPO, an open-source, RDKit-based Python package for generating IUPAC names, developed by an agentic self-improvement loop using Codex.
Development took 40 days, with the implementation agent active on 29 of these, and consumed approximately 20 billion tokens. 
On our held-out benchmark of 103 million PubChem molecules, NISPO achieved 98.1\% round-trip accuracy. 

We defined the accuracy of NISPO by round-trip success against OPSIN. 
Consequently, while NISPO names are valid, they are unlikely to match the ``preferred'' names defined in the IUPAC Blue Book. 
For this reason, NISPO is not suitable for cases where generated names must follow strict IUPAC nomenclature rules.

The NISPO codebase does not conform to conventional human standards for code quality. 
For example, one script in NISPO contains 22,203 lines. 
A code-quality objective might have guided the agent towards a more maintainable implementation, or a refactoring phase could be used to restructure the codebase. 
For our use case, the human readability of the codebase was not a property we required, so we did not optimise nor assess for this. 

We have demonstrated that agentic self-improvement loops can autonomously develop scientific software when progress can be measured against a verifiable objective.
We expect this method can be used to develop software tools across numerous scientific domains where such objectives exist, and, where experimental outcomes can themselves serve as verifiers, may extend towards autonomous scientific discovery.
We hope NISPO proves useful as a tool for the chemistry community, and that its development strategy inspires further agent-developed scientific software.

\section*{Software and Data Availability}
The NISPO tool is available at \mbox{\url{https://github.com/oxpig/nispo}}, and can be installed from the Python Package Index using \texttt{pip install nispo}.

\section*{Acknowledgments}
N.T.R. is supported by UK Research and Innovation (UKRI) through an EPSRC-funded PhD studentship (project reference: 2928891).
We acknowledge OpenAI for providing a ChatGPT Pro subscription used in this study for running Codex. 
The authors would like to acknowledge the use of the University of Oxford Advanced Research Computing (ARC) facility in carrying out this work.

\section*{Declaration of AI use}
NISPO was developed using OpenAI's Codex with the GPT-5.5 model. Claude Fable 5 was used to provide feedback on the writing of this paper.

\bibliographystyle{plainnat6doiElseURL}
\bibliography{references}

\clearpage

\appendix

\section{Original AGENTS.md}

\begin{lstlisting}[caption={Original AGENTS.md file used for development of NISPO. During the full optimisation process the AGENTS.md file was refined, such as by adding additional objectives, analysis helpers, and integrity constraints.},label={lst:original-agents-file}]
# NISPO Project Instructions

## Objective

NISPO is a standalone Python package for converting SMILES strings to IUPAC-style names. Its primary target is OPSIN roundtrip success:

```text
SMILES -> NISPO name -> OPSIN -> SMILES
```

The benchmark score is the fraction of molecules where the input SMILES and the OPSIN-parsed SMILES are equivalent after RDKit canonicalization. The target is 98% roundtrip accuracy on the benchmark in `benchmark/`.

Optimize for valid OPSIN-roundtripping names first. Preferred or especially pretty IUPAC names are secondary.

## Required NISPO API

The package in `nispo/` must expose:

```python
from nispo import smiles_to_iupac, smiles_to_iupac_batch
```
- `smiles_to_iupac(smiles: str) -> str` is the single-molecule debugging API.
- `smiles_to_iupac_batch(smiles_list) -> list[str]` is the benchmark API.

The batch API must return exactly one item per input SMILES. For a molecule that cannot be named, return an empty string rather than aborting the whole batch.

## Benchmark

Run the full benchmark with:

```bash
conda run -n rdkit python benchmark/score_roundtrip.py
```

The benchmark scorer reads the SMILES inputs from:

```text
benchmark/benchmark_roundtrip_smiles.txt
```

Each run writes per-molecule results under `benchmark/runs/` and appends a summary to `benchmark/history.jsonl`.

Do not add alternate scoring paths unless the user explicitly asks for them. The standard score should remain one simple full-benchmark command.

## Sources Of Truth

Use these sources in this order:

1. `opsin/`: primary behavioral reference. Study OPSIN parser rules, resources, tokenization, suffix handling, locants, nomenclature classes, and failure behavior. NISPO should be an independent inverse rule system, but OPSIN is the oracle for what names will be accepted.
2. `iupac_blue_book/`: supporting IUPAC rule context, especially when OPSIN behavior needs interpretation or a rule needs a canonical formulation.
3. RDKit: molecular graph parsing, canonicalization, ring perception, aromaticity handling, and functional-group inspection.

NISPO must not depend on OPSIN internals at runtime. OPSIN is only the scoring oracle and rule reference.

## Anti-Cheating Rules

Do not use external answer files or previously generated IUPAC names to name benchmark molecules.

Do not create benchmark-specific lookup tables, memorize individual benchmark rows, or add one-off special cases that are not general nomenclature rules.

Do not modify the benchmark SMILES list, scorer, OPSIN jar, or scoring semantics to improve the score unless the user explicitly approves the change.

Implement general naming rules in NISPO. Benchmark improvements should come from better chemistry and nomenclature coverage, not from row-specific optimization.

## Work Strategy

Goal-mode agents should create their own subgoals and keep implementing until the tool is complete or genuinely blocked.

The expected loop is:

1. Run `benchmark/score_roundtrip.py` to measure current performance.
2. Inspect failure classes in the latest `benchmark/latest/results.tsv`.
3. Study relevant OPSIN rules/resources and Blue Book sections.
4. Implement general NISPO naming rules.
5. Rerun the full benchmark and compare `benchmark/history.jsonl`.
6. Repeat, prioritizing changes that improve broad classes of molecules.

Preserve batch performance. Both NISPO generation and OPSIN scoring should remain batch-oriented.
\end{lstlisting}

\clearpage
\section{Goal prompt}

\begin{lstlisting}[caption={Example goal prompt used for initial implementation of NISPO. Prompt was designed by a Codex GPT-5.5 agent following discussion with the human researcher.},label={lst:goal-prompt}]
/goal Implement NISPO in /Users/nicholas/projects/iupac_naming/nispo as a standalone Python package that converts SMILES strings to IUPAC-style names.

Use /Users/nicholas/projects/iupac_naming/AGENTS.md as the project contract. NISPO must expose:

from nispo import smiles_to_iupac, smiles_to_iupac_batch

Benchmark with:

conda run -n rdkit python benchmark/score_roundtrip.py

Work autonomously toward at least 99% OPSIN roundtrip accuracy on the benchmark in /Users/nicholas/projects/iupac_naming/benchmark. Use opsin/ as the primary behavioral reference, iupac_blue_book/ as supporting nomenclature context, and RDKit for molecular graph handling.

Iterate independently: run the benchmark, inspect benchmark/latest/results.tsv, study the relevant OPSIN/IUPAC rules, implement general deterministic naming rules in NISPO, rerun, and continue improving until the score reaches 0.99 or the work is genuinely blocked.

Do not use external answer files, previously generated IUPAC names, benchmark-specific lookup tables, row-specific special cases, or scorer/benchmark modifications to improve the score. NISPO must remain independent of OPSIN at runtime; OPSIN is only the validation oracle and rule reference.
\end{lstlisting}

\clearpage

\section{NISPO analysis suite}

\begin{lstlisting}[caption={NISPO analysis suite skill. This skill defined a set of analysis commands that the Codex agent could use to interpret the results of benchmark runs.},label={lst:nispo-analysis-suite}]
---
name: nispo-analysis-suite
description: Analyze NISPO benchmark runs: summaries, failure families, telemetry bottlenecks, deltas, and implementation-location hints. Analysis only; fixes must be graph/rule/renderer mechanisms, not string rewrites.
---

# NISPO Analysis Suite

Router command:

```bash
conda run -n rdkit python /Users/nicholas/.codex/skills/nispo-analysis-suite/scripts/nispo_analysis.py <command> [args]
```

Use `<command> --help` for options.

Commands:
- `latest-run`: latest fixed-benchmark metrics; optional previous-run delta.
- `paired-names`: diagnostic table joining generated names, statuses, SMILES, source metadata, and row ids; use to inspect failure families, not copy source names.
- `generated-patterns`: generated-name diagnostics for locating renderer/rule family problems; do not convert string patterns into runtime policy.
- `long-tail-patterns`: long-name diagnostics for identifying renderer/parent-choice issues.
- `implementation-map`: non-binding hints for likely NISPO core subsystem.
- `benchmark-delta`: compare two run summaries.
- `backlog`: optional local candidate notes.

Graph-derived class tooling lives in the repo, not this router:
```bash
conda run -n rdkit python benchmark/analysis/nispo_analysis_suite/molecule_classes.py <command> [args]
```

Commands:

- `class-summary`: summarize benchmark status by graph-derived molecule class.
- `subset-by-class`: create benchmark-compatible subsets by class/status, preserving row mapping.
- `classify-set`: write per-set `molecule_classes.tsv` if missing.

Use these before broad optimization work to target large weak classes. Class labels are graph/RDKit-derived only; do not convert class diagnostics into row-specific fixes or source/generated-name string policy.

Run `paired-names` without `--limit` after each new full benchmark before using `generated-patterns` or `long-tail-patterns`. Limited paired outputs are written to separate files and are not reused as canonical pattern inputs.

Defaults: project root `/Users/nicholas/projects/iupac_naming`; set `benchmark/sets/016_cluster_0045_failures_remaining_clean_organic`; outputs `benchmark/analysis/nispo_analysis_suite/`.

Row indexes: prefer explicit metadata columns such as `benchmark_row_index_0based`, `benchmark_output_row_index_1based`, and `source_row_0based`. Local benchmark outputs keep 1-based `row_index`. Analysis outputs should use explicit `*_0based`/`*_1based` names.

Guardrails: analysis only. Use benchmark strings only as examples attached to graph/status/failure evidence. Do not convert string patterns, generated substrings, or example names into runtime lookup tables, final-output rewrites, rendered-name selector policy, or benchmark-specific shortcuts. Implementation must use graph/candidate facts, provenance, renderer capability, nomenclature resources, or rule state.


\end{lstlisting}

\clearpage

\begin{figure}
    \centering
    \includegraphics[width=0.75\linewidth]{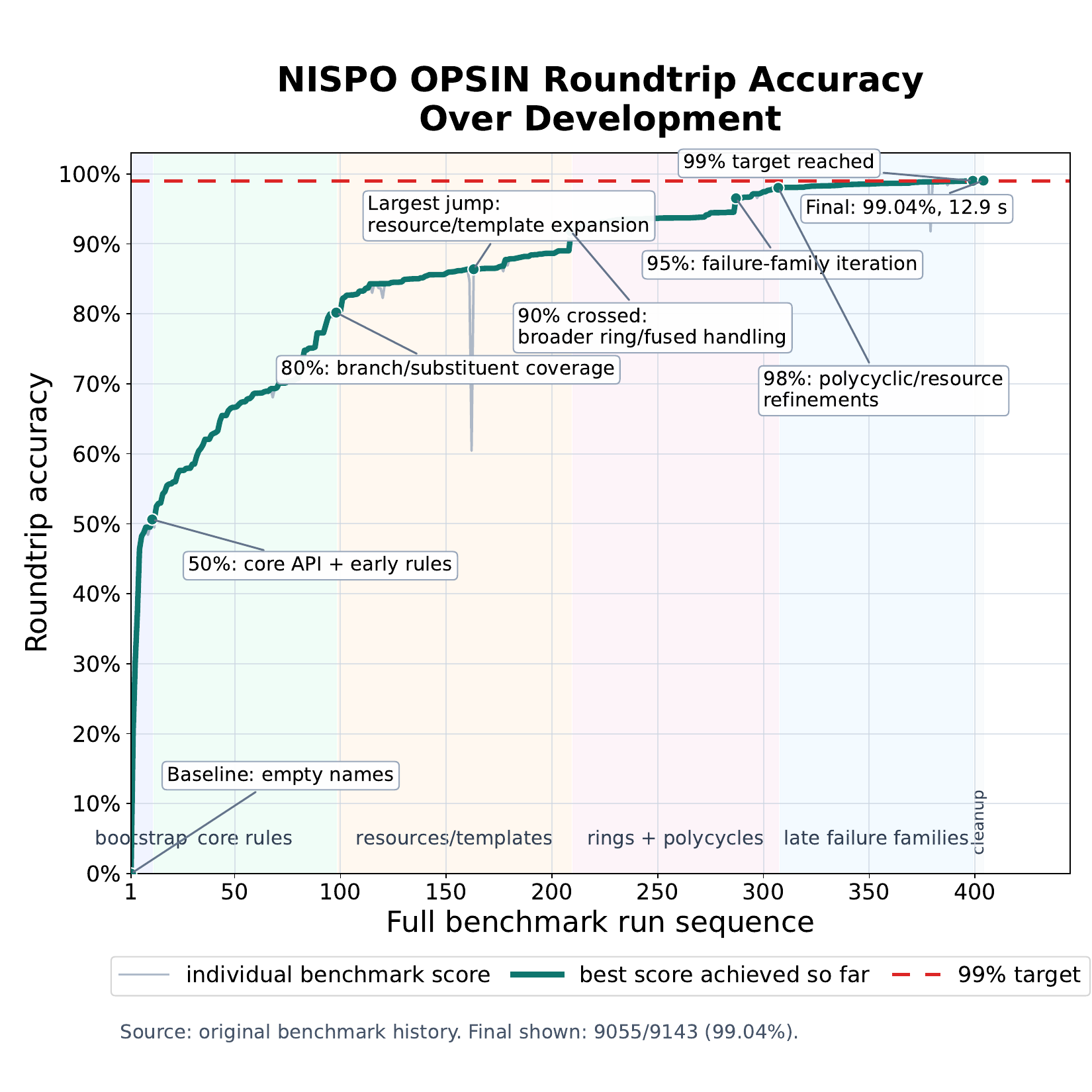}
    \caption{Improvement of NISPO on the MolJSON benchmark over time. The result of each full benchmark run triggered by the implementation agent is plotted. The Codex GPT-5.5 agent was prompted to annotate key implementation stages during the development process, and to generate this figure.}
    \label{fig:learning_graph_on_moljson}
\end{figure}

\clearpage

\clearpage

\section{Final AGENTS.md}

\begin{lstlisting}[caption={Final AGENTS.md file from NISPO project. During the project the AGENTS.md file was updated to alter the behavior of the implementation agent.},label={lst:final-agents-file}]
# NISPO Instructions

NISPO: SMILES -> IUPAC-style name. Correctness = deterministic OPSIN
roundtrip: `SMILES -> NISPO name -> OPSIN -> SMILES`.

## Paths

- Root: `/Users/nicholas/projects/iupac_naming`
- NISPO repo/tool: `/Users/nicholas/projects/iupac_naming/nispo`
- Benchmarks: `benchmark/sets/`
- IUPAC refs: `iupac_blue_book/`
- OPSIN source/ref: `opsin/`

Run benchmarks from root. Edit package code at `nispo/nispo/...` from root, or
`nispo/...` inside the NISPO repo.

## API

Preserve `from nispo import smiles_to_iupac, smiles_to_iupac_batch`.
Batch output count == input count. Unsupported/failures return `""`; batches
must not abort.

## Benchmark

Default target: `benchmark/sets/016_cluster_0045_failures_remaining_clean_organic`

Authoritative local run:

```bash
conda run -n rdkit python benchmark/run_benchmark.py benchmark/sets/016_cluster_0045_failures_remaining_clean_organic --workers 8 --chunk-size 1000 --opsin-shard-size 1000 --progress-molecules 0
```

Use explicit chunk sizes to avoid long-tail wall-clock stalls from coarse shards.

Local subsets allowed; preserve source row mapping. Broader parent:
`benchmark/sets/015_cluster_0045_failures_remaining`.

Run outputs: `benchmark/sets/<set>/runs/<timestamp>/`, `latest`,
`history.jsonl`. Inspect `summary.json`, `row_status.tsv`,
`generated_names.tsv`, `telemetry.json`, `name_length_comparison.tsv`.

Reference-length metric:
`summary["name_length_vs_surechembl"]["match_nispo_shorter_or_same_rate"]`.

Indexes: prefer explicit metadata columns such as `benchmark_row_index_0based`,
`benchmark_output_row_index_1based`, and `source_row_0based`; local run outputs
use 1-based `row_index`.

Cluster is disabled for now. If the user asks for cluster work, first restore
`.codex/skills/nispo-cluster-benchmark/SKILL.disabled.md` to `SKILL.md`.

## Optimization

For NISPO optimization goals, start with `docs/nispo_optimization_context/00_start_here.md`; open heavier docs/Graphify only on demand.

- Treat NISPO core as greenfield when needed: prefer clean graph-first
  rewrites over preserving legacy internals, while keeping public API and
  benchmark contracts stable.
- Broad chemistry mechanisms, not rows/source-name patterns/selector knobs.
- Prefer graph-first: shared descriptors, structured candidates, renderer capability, provenance.
- Ranking/selection must use graph/candidate facts, never rendered-name spelling/regex/substrings.
- Empty outputs = coverage failures first; use telemetry for missing candidates/renderers/parents/suffixes/repeated work.
- Validate mixed samples; track recoveries/regressions/parse failures/timing/reference length.

## Validation

Use OPSIN for roundtrip, RDKit for graph equivalence. Batch OPSIN. Preserve
OPSIN row/name alignment; `-n` output is `SMILES<TAB>name`. Compare isomeric
canonical SMILES. SureChEMBL names may inform analysis only; never runtime
lookup/answers.
OPSIN Java: prefer `/opt/homebrew/opt/openjdk/bin/java`; `/usr/bin/java` may be
the macOS stub. For probes, set `JAVA=/opt/homebrew/opt/openjdk/bin/java` or put
`/opt/homebrew/opt/openjdk/bin` first in `PATH`.

## Integrity

No row-specific fixes, answer/source-name lookup, memorization tables, scorer
changes, runtime SureChEMBL lookup, exact benchmark shortcuts, final-output
compaction, downstream semantic regex/string rewrites, or rendered-name spelling
policy in candidate selection.

Short names must come from rules/renderers/parent choice/templates/graph-first
candidate facts/structured ranking. General nomenclature tables/templates
allowed; final rewrite maps are not. Runtime deterministic.

## Tests/Git

Tests: prefer API, determinism, OPSIN roundtrip, graph equivalence, telemetry.
Avoid exact-string/internal-lock tests.

Git: commit coherent tested NISPO progress regularly after validated mechanisms
or tooling/diagnostic improvements. No failed experiments, bulky runs,
unrelated changes, or push unless asked.

## graphify

This project has a knowledge graph at graphify-out/ with god nodes, community structure, and cross-file relationships.

When the user types `/graphify`, invoke the `skill` tool with `skill: "graphify"` before doing anything else.

Rules:
- For codebase questions, first run `graphify query "<question>"` when graphify-out/graph.json exists. Use `graphify path "<A>" "<B>"` for relationships and `graphify explain "<concept>"` for focused concepts. These return a scoped subgraph, usually much smaller than GRAPH_REPORT.md or raw grep output.
- Dirty graphify-out/ files are expected after hooks or incremental updates; dirty graph files are not a reason to skip graphify. Only skip graphify if the task is about stale or incorrect graph output, or the user explicitly says not to use it.
- If graphify-out/wiki/index.md exists, use it for broad navigation instead of raw source browsing.
- Read graphify-out/GRAPH_REPORT.md only for broad architecture review or when query/path/explain do not surface enough context.
- After modifying code, run `graphify update .` to keep the graph current (AST-only, no API cost).

\end{lstlisting}

\end{document}